%% file: main.tex
\documentclass[3p,times,11pt]{elsarticle}
\usepackage{lineno}

\let\today\relax
\makeatletter
\def\ps@pprintTitle{%
    \let\@oddhead\@empty
    \let\@evenhead\@empty
    \def\@oddfoot{\footnotesize\itshape
         {} \hfill\today}%
    \let\@evenfoot\@oddfoot
    }
\makeatother

\input{listofpackages.tex}

\input{definitions.tex}
\DeclareMathAlphabet{\pazocal}{OMS}{zplm}{m}{n}
\newcommand{\Q}{\pazocal{Q}}

\journal{ESPPU}

\begin{document}

\begin{frontmatter}

\title{Physics with high-luminosity proton-nucleus collisions at the LHC\tnoteref{t1}\vspace*{-0.25cm}}

\tnotetext[t1]{This is a slightly expanded version of a contribution submitted for the 2025 European Strategy for Particle Physics. Its original version can be found at \url{https://indico.cern.ch/e/pA4LHC} along with the list of the 150+ people who endorsed it.}

\include{authors.tex}

\begin{abstract}
The physics case for the operation of high-luminosity proton-nucleus (\pA) collisions during Run 3 and 4 at the LHC is reviewed. The collection of $\mathcal{O}$(1--10~pb$^{-1}$) of proton-lead (\pPb) collisions at the LHC will provide unique physics opportunities in a broad range of topics including proton and nuclear parton distribution functions (PDFs and nPDFs), generalised parton distributions (GPDs), transverse momentum dependent PDFs (TMDs), low-$x$ QCD and parton saturation, hadron spectroscopy, baseline studies for quark-gluon plasma and parton collectivity, double and triple parton scatterings (DPS/TPS), photon-photon collisions, and physics beyond the Standard Model (BSM); which are not otherwise as clearly accessible by exploiting data from any other colliding system at the LHC. This report summarises the accelerator aspects of high-luminosity \pA\ operation at the LHC, as well as each of the physics topics outlined above, including the relevant experimental measurements that motivate --much-- larger \pA\ datasets.
\end{abstract}

\end{frontmatter}

\input{introduction.tex}

\input{accelerator.tex}

\input{intro_physics_case.tex}

\input{nPDF.tex}

\input{1photon.tex}

\input{saturation.tex}

\input{AA-benchmark.tex}

\input{DPS-TPS.tex}

\input{spectroscopy}

\input{2photon.tex}

\input{BSM.tex}

\input{UHECR.tex}

\input{summary-conclusions.tex}

\input{acknowledgements.tex}


\clearpage
\bibliographystyle{myutphys}
\bibliography{references}%

\end{document}

%% file: listofpackages.tex
\usepackage{xspace}
\usepackage{lineno}
\usepackage{graphicx}%
\usepackage{multirow}%
\usepackage{amsmath,amssymb,amsfonts}%
\usepackage{amsthm}%
\usepackage{mathrsfs}%
\usepackage[title]{appendix}%
\usepackage[table,xcdraw]{xcolor}
\usepackage{textcomp}%
\usepackage{manyfoot}%
\usepackage{booktabs}%
\usepackage{algorithm}%
\usepackage{algorithmicx}%
\usepackage{algpseudocode}%
\usepackage{listings}%
\usepackage{placeins}
\usepackage{subfig}
\usepackage{subfiles}%
\usepackage{textgreek}
\usepackage{comment}
\usepackage{siunitx}
\usepackage{tikz} 
\usepackage{tabularx}       
\usepackage{makecell}       
\usepackage{pdflscape} 
\usepackage{booktabs} 
\usepackage{amsmath} 
\usepackage{makecell}
\usepackage{multirow} 
\usepackage{tcolorbox} 
\usepackage[normalem]{ulem}
\usepackage{cancel}
\usepackage{wrapfig}
\usepackage{rotating}
\usepackage{wasysym}   
\usepackage{hyperref}
\hypersetup{%
        colorlinks,
        breaklinks=true,
        plainpages=false,%
        citecolor=blue,
        linkcolor=blue,
        urlcolor=blue,
        bookmarksopen=true,%
        bookmarksnumbered=false,%
        bookmarksdepth=5%
}

\usepackage{bbding}
\usepackage{pifont}

%% file: definitions.tex
\newcommand{\pt}{\ensuremath{p_\mathrm{T}}\xspace}
\newcommand{\pT}{\ensuremath{p_\mathrm{T}}\xspace}
\newcommand{\kt}{\ensuremath{k_\mathrm{T}}\xspace}

\newcommand{\micron}{\ensuremath{\upmu \mathrm{m}}}

\newcommand{\keV}{\si{\kilo\electronvolt}}
\newcommand{\MeV}{\si{\mega\electronvolt}}
\newcommand{\GeV}{\si{\giga\electronvolt}}

\newcommand{\mlab}[1]%
    {\mbox{}\marginpar{\raggedright\hspace{0pt}\tiny {\color{red}{#1}}}}

\newcommand{\as}{\alpha_{\rm S}}

\newcommand{\sqrts}{\sqrt{s}}

\newcommand{\invab}{{\rm ab}^{-1}}

\newcommand{\LumiInt}{\mathcal{L}_\text{int}}

\newcommand{\PbPb}{PbPb}

\newcommand{\pPb}{$p$Pb}
\newcommand{\pA}{$pA$}
\renewcommand{\AA}{$AA$}
\newcommand{\pp}{$pp$}

\newcommand {\fmoverc}			{\mbox{\rm fm$\kern-0.15em /\kern-0.12em c$}}
\newcommand {\etaovers}			{\mbox{\rm $\eta \kern-0.05em/\kern-0.15em s$}}
\newcommand {\gevoverfm}		{\mbox{\rm GeV$\kern-0.15em /\kern-0.10em \mathrm{fm}^3$}}

\newcommand{\sqrtsnn}{\!\sqrt{s_{NN}}}

\definecolor{verylightgray}{gray}{0.95}
\definecolor{midgray}{gray}{0.7}
\definecolor{blue1}{rgb}{0.1875000, 0.320312, 0.60546875}
\definecolor{blue2}{rgb}{0.4960937, 0.640625, 0.82421875}
\definecolor{blue3}{rgb}{0.6445312, 0.777344, 0.90625000}



%% file: authors.tex
\author[CERN]{\small D.~d'Enterria\fnref{Editor}}

\address[CERN]{CERN, EP Department, 1211 Geneva, Switzerland}

\author[IJCLAB]{\small C.~A.~Flett\fnref{Editor}}

\address[IJCLAB]{Université Paris-Saclay, CNRS, IJCLab, 91405 Orsay, France}

\author[AGH]{\small I.~Grabowska-Bold\fnref{Editor}}

\address[AGH]{AGH University of Krakow, Faculty of Physics and Applied Computer Science, Krakow}

\author[IJCLAB]{\small C.~Hadjidakis\fnref{Editor}}

\author[AGH]{\small P.~Kotko\fnref{Editor}}

\author[IFJ]{\small A.~Kusina\fnref{Editor}}

\address[IFJ]{Institute of Nuclear Physics, Polish Academy of Sciences, PL-31342 Krakow, Poland}

\author[IJCLAB]{\small J.P.~Lansberg\fnref{Editor}}

\author[UCD]{\small R.~McNulty\fnref{Editor}}

\address[UCD]{School of Physics, University College Dublin, Dublin 4, Ireland}

\author[PER]{\small M.~Rinaldi\fnref{Editor}}

\address[PER]{Dipartimento di Fisica.~Università degli studi di Perugia, and INFN Sezione di Perugia.~Via A.~Pascoli, Perugia, 06123, Italy}

\author[FIR]{\small L.~Bonechi\fnref{Contributor}}
\address[FIR]{INFN, Sezione di Firenze, Via B.~Rossi 1, Sesto Fiorentino 50019, Italy}

\author[CERN2]{\small R.~Bruce\fnref{Contributor}}
\address[CERN2]{CERN, BE Department, 1211 Geneva, Switzerland}

\author[ALA]{\small C.~Da~Silva\fnref{Contributor}}
\address[ALA]{Los Alamos National Laboratory (LANL), Los Alamos, NM, United States}

\author[SAN]{\small E.G.~Ferreiro\fnref{Contributor}}
\address[SAN]{Instituto Galego de Física de Altas Enerxías, Univ.\ Santiago de Compostela, E-15782, Galicia, Spain}

\author[SP]{\small S.~Fichet\fnref{Contributor}}
\address[SP]{CCNH, Universidade Federal do ABC, Santo André, 09210-580 SP, Brazil}

\author[UCL]{\small L.~Harland-Lang\fnref{Contributor}}
\address[UCL]{Department of Physics and Astronomy, University College London, London, WC1E 6BT, UK}

\author[MIT]{\small G.~Innocenti\fnref{Contributor}}
\address[MIT]{Massachusetts Institute of Technology, 371 Richmond Blvd, Ronkonkoma, NY 11779, USA}

\author[BER]{\small F.~Jonas\fnref{Contributor}}
\address[BER]{Lawrence Berkeley National Laboratory, Berkeley, California, United States}

\author[GSI]{\small J.~M.~Jowett\fnref{Contributor}}
\address[GSI]{GSI Helmholtzzentrum für Schwerionenforschung GmbH, Darmstadt, Germany}

\author[ILL]{\small R.~Longo\fnref{Contributor}}
\address[ILL]{Department of Physics, University of Illinois at Urbana-Champaign, Urbana IL, USA}

\author[IJCLAB,UCD]{\small K.~Lynch\fnref{Contributor}}

\author[MIT]{\small C.~McGinn\fnref{Contributor}}

\author[KAR]{\small T.~Pierog\fnref{Contributor}}
\address[KAR]{Karlsruhe Institute of Technology
76131 Karlsruhe, Germany}

\author[CERN]{\small M.~Pitt\fnref{Contributor}}

\author[CERN2]{\small S.~Redaelli\fnref{Contributor}}

\author[BRO]{\small B.~Schenke\fnref{Contributor}}
\address[BRO]{Physics Department, Brookhaven National Laboratory, Upton, NY 11973, USA}

\author[GRE]{\small I.~Schienbein\fnref{Contributor}}
\address[GRE]{Lab.\ Physique Subatomique et Cosmologie, Univ.\ Grenoble-Alpes, CNRS/IN2P3, 53 Av.\ des Martyrs, 38026 Grenoble, France} 

\author[OHI]{\small M.~Stefaniak\fnref{Contributor}}
\address[OHI]{The Ohio State University, Department of Physics, 191 Woodruff Avenue, Columbus, OH 43210}

\author[PEN]{\small M.~Strikman\fnref{Contributor}}
\address[PEN]{Penn State University, University Park, PA 16802, PA, USA}

\author[WAR]{\small L.~Szymanowski\fnref{Contributor}}
\address[WAR]{National Centre for Nuclear Research (NCBJ), Pasteura 7, 02-093 Warsaw,
Poland}

\author[KAN]{\small D.~Tapia~Takaki\fnref{Contributor}}
\address[KAN]{Department of Physics and Astronomy, The University of Kansas, Lawrence, KS, 66045 USA}

\author[ALC]{\small C.~Van~Hulse\fnref{Contributor}}
\address[ALC]{Universidad de Alcalá, Alcalá de Henares , Spain\vspace*{-1cm}}

\author[IJCLAB]{\small S.~Wallon\fnref{Contributor}}

\fntext[Editor]{Editor}
\fntext[Contributor]{Contributor}

%% file: introduction.tex
\section{Introduction}
\label{sec:Introduction}

This document reports the physics case for an ambitious proton-nucleus (\pA) collision programme at the LHC in the context of the forthcoming update of the European strategy for particle physics. Following the physics case outlined in Ref.~\cite{Salgado:2011wc}, and a short pilot run to demonstrate feasibility in 2012, the LHC operated \pPb\ collisions at nucleon-nucleon ($NN$) center-of-mass (CM) energies of $\sqrtsnn = 5.02,\,8.16$~TeV in 2013 and 2016, respectively, but no run has been performed since then, and none is currently planned for the near future. The past \pPb\ runs have brought essential contributions to particle, heavy-ion, and cosmic-ray physics, leading to, among others, significantly improved nuclear parton distributions functions (nPDFs)~\cite{Klasen:2023uqj}, and the discovery of new phenomena, such as the onset of parton collectivity~\cite{Grosse-Oetringhaus:2024bwr}. As discussed hereafter, high-luminosity \pA\ collisions in Run~3~and~4 at $\sqrtsnn = 8.54$~TeV, involving both heavy and light nuclei, are essential to fully exploit a rich experimental programme for the study of quantum chromodynamics (QCD) in the perturbative, nonperturbative, and high-density regimes. They provide, in particular, a unique and complementary environment for uncovering the tomography of the proton and nuclei, and their partonic properties.

Compared to \pp\ collisions, \pA\ interactions allow the exploration of nuclear modifications to PDFs and cold nuclear matter effects, which are essential for understanding the initial-state conditions of heavy-ion collisions. Additionally, \pA\ collisions probe small-$x$ physics more effectively than \pp, providing insight into gluon saturation and the onset of nonlinear QCD effects, which are enhanced due to the intrinsically larger number of initial partons~\cite{Jalilian-Marian:2005ccm}.

Compared to \AA\ collisions, \pA\ collisions offer higher nucleon-nucleon CM energies, higher luminosity, and a cleaner environment, free from the final-state complexities of a fully-developed QGP, making them an essential benchmark for interpreting heavy-ion data, while studying the onset of collectivity in small systems. Photoproduction in \pA\ collisions, which capitalises on the LHC as a photon-collider, complements the ultraperipheral-collisions (UPCs) programme in \AA\ collisions~\cite{Baltz:2007kq}, and has many benefits through being able to distinguish the more energetic photon emitter.  

Investigations of quarkonium production~\cite{Chapon:2020heu}, exotic hadrons, scenarios beyond the Standard Model (BSM), and hadronisation mechanisms, can all be performed in detail in an environment that bridges the gap between \pp\ and \AA\ systems, and improves our understanding of ultrahigh-energy cosmic-ray interactions~\cite{dEnterria:2011twh}, underscoring the importance of \pA\ data in advancing our knowledge of high-energy nuclear, particle, and astroparticle physics. The data taken in \pA\ collisions at Run 1 and 2 of the LHC programme has had a major impact for all these fields, extrapolating precise knowledge of proton collisions to multi-nucleon systems.
The novel use of these collisions has brought significant advances to our understanding of both the proton and the nucleus. The Run 3 plan discussions have focused so far on \pp\ and \AA\ collisions: the absence of \pA\ collisions to date is notable and appears as a missed opportunity. This document presents a summary of the accelerator and physics case aspects needed to revert this trend.

%% file: accelerator.tex
\section{Accelerator and detector considerations}
\label{acc}

The LHC has been designed to collide protons and nuclei at a beam energy of $7\,Z$~TeV (where $Z$ is the ion electric charge)~\cite{lhcdesignV1} and up to $6.8\,Z$~TeV has been achieved to date. The LHC typically operates about one month per year with heavy-ion beams, mainly fully stripped Pb nuclei. Initially, the heavy-ion programme consisted only of \PbPb\ collisions, and it was then extended with a new mode of operation with proton-nucleus collisions~\cite{Jowett:2006au,Salgado:2011wc,Jowett:2013uka, Versteegen:2013cza,Jowett:2017dqj, Jebramcik2019}. Following initial pilot tests, two 1-month physics runs with \pPb\ collisions were carried out in 2013 and 2016, with integrated luminosities of $\LumiInt \approx 220$~nb$^{-1}$ collected in ATLAS and CMS, 75~nb$^{-1}$ in ALICE, and 36~nb$^{-1}$ in LHCb, combining the data in Run~1 (2010--2013) and Run~2 (2015--2018). The heavy-ion operation in the ongoing Run~3 (scheduled from 2022 to mid 2026) has consisted of two \PbPb\ runs so far. In the future, collisions with Pb ions are scheduled to continue with 1-month heavy-ion operation in most operational years until the end of Run~4 (scheduled for 2030--2033). Operation with \pPb\ is included in this plan, but the detailed \PbPb\ and \pPb\ time allocations have not yet been decided. The goals for future \pPb\ operation, combining Run~3 and Run~4, are $\LumiInt=1.2$~pb$^{-1}$ at ATLAS and CMS, and 0.6~pb$^{-1}$ at ALICE and LHCb~\cite{Citron:2018lsq}. The next opportunity for \pPb\ operation might come already in 2026, however, the decision has not yet been taken. In addition to high-intensity \pPb\ operation, a short low-intensity $p$O run is planned for mid-2025~\cite{Bruce2021a}. 

The assumed LHC scenario for future \pPb\ operation considers the same machine cycle as for \PbPb~\cite{bruce20_HL_ion_report}, relying on crystal collimation~\cite{DAndrea:2024sui}, and round optics with $\beta^*=0.5$~m at ALICE, ATLAS, and CMS. However, a more complicated setup of the radiofrequency (RF) system is needed. Because of the difference in charge-to-mass ratio between protons and Pb, the two species have different revolution frequencies at equal momentum per charge. Therefore, both beams have to be brought off-momentum in different directions by the RF cavities 
to equalise their 
frequencies, such that the longitudinal locations of the collision points are stationary. This momentum offset is introduced only at top energy due to aperture constraints. An additional challenge is the beam-beam effect between the asymmetric beams and the moving long-range beam-beam encounters~\cite{Jebramcik2019,Jebramcik:2019mro}. We assume the same structure of the Pb beam as in \PbPb\ operation, with 50-ns bunch spacing by interleaving different bunch trains longitudinally in the SPS (``slip-stacking''), and an opposing 50-ns proton beam with low intensity. The proton beam is produced in a different way by the injectors, and a perfect overlap between the two beams cannot be obtained, resulting in slightly fewer colliding bunches per experiment. It is also assumed that only ALICE needs luminosity levelling at $5\times10^{29}$\,cm$^{-2}$s$^{-1}$ in order to limit the event rate to about 1~MHz. 

The projected future luminosity performance in single \pPb\ fills have been simulated with the 
CTE code~\cite{bruce10prstabCTE} 
extrapolated over a full 1-month run~\cite{Bruce:2021hii}. Recent calculations with updated filling schemes give projected $\LumiInt \approx 0.33$~pb$^{-1}$ at ALICE, 0.47~pb$^{-1}$ at ATLAS/CMS, 0.15~pb$^{-1}$ at LHCb, for a 1-month run (24 days of physics) at $\sqrtsnn=$~8.54~TeV (6.8~$Z$~TeV beam energy). These numbers carry large uncertainties and depend highly on machine availability and achieved beam parameters. If \pPb\ collisions are performed at the lower \PbPb\ beam energy, to be used as reference data for the latter and profiting from the same \pp\ reference data set, the luminosity would be reduced. Table~\ref{tab:lumipPbRun34} summarises the delivered $\LumiInt$, the targets for future runs, the projected $\LumiInt$ per run, and the number of runs needed to reach the targets. We note that one day of high-luminosity \pPb\ running corresponds to 10--20\% of the total $\LumiInt$ gathered thus far. Given that up to four 1-month runs are needed to reach the target $\LumiInt$, and there will likely be at most two \pPb\ runs until the end of Run~4, several ways of improving the performance are being explored. 

\begin{table}[h!]
\tabcolsep=4.5mm
\caption{
Delivered integrated luminosities in \pPb\ collisions in Run~1 and Run~2, $\LumiInt$ targets in \pPb\ collisions in Run~3 and 4, and number of runs needed to achieve these $\LumiInt$ with the presently predicted performance. 
\label{tab:lumipPbRun34}}
\centering
\begin{tabular}{p{8.0cm}|c|c|c} 
& ALICE & ATLAS/CMS & LHCb\\
 \hline
Total $\LumiInt$ delivered in Run~1 and Run~2 (nb$^{-1}$) & 75 & 220 & 36 \\
\hline
Target $\LumiInt$ for Run~3 and Run~4 (pb$^{-1}$) & 0.6 & 1.2 & 0.6 \\
Projected $\LumiInt$ per 1-month run (pb$^{-1}$)~\cite{RoderickWorkshop} 
 & 0.33 & 0.47 & 0.15\\
Number of runs needed to reach targets & 1.8 & 2.5 & 4
 \\ \hline
\end{tabular}
\end{table}

Increasing the Pb intensity, which might be within reach based on the injector performance in 2024 and on the upgrades deployed at the LHC~\cite{Arduini:2024kdp}, is a good way forward, but likely not enough to reach the targets at all IPs in two runs. Decreasing the proton-beam bunch spacing to 25~ns, as used in standard \pp\ operation, allows for more collisions at LHCb without penalizing other experiments. The peak luminosity can be further increased with higher proton bunch intensity, which would benefit all experiments except ALICE, assumed to be levelling. 
Further improvements might come from reduced $\beta^*$ and crossing angles. However, studies are needed to investigate the feasibility of all these measures, e.g.\ in view of the much stronger beam-beam effects, as well as beam instrumentation in case of very asymmetric beams. It therefore remains as future work to investigate realistic scenarios where all experiments meet their $\LumiInt$ targets. 

In Run 4, the \pPb\ programme will benefit from 
improved experimental setups, including enhanced forward detection capabilities (e.g., extended ATLAS and CMS trackers over $|\eta| < 4$~\cite{ATL-PHYS-PUB-2021-024, Rossi:2816248} and ALICE FoCal with $3.2<\eta<5.8$~\cite{CERN-LHCC-2024-004}), and forward proton detectors (e.g., CMS PPS~\cite{Collaboration:2750358}). Beyond Run~4, ion operation at the LHC is planned to continue, including the planned ALICE~3 detector~\cite{alicecollaboration2022letterintentalice3} for Run 5 and beyond.
The main goal for this period will be to produce significantly higher nucleon-nucleon luminosities, and other nuclei than Pb are being investigated. Studies of achievable intensities for a range of ion species are ongoing in the CERN injector complex. A detailed programme for this period has not been elaborated yet, and \pA\ collisions might be included. Initial studies of the foreseen performance have been presented in~\cite{Jebramcik2019,Jebramcik:IPAC2019-MOPMP024}, although the achievable ion bunch intensities will need to be revised in the future. Furthermore, as there are no technical limitations to colliding protons with other nuclei than Pb, short low-intensity runs in such configurations may be envisaged, similarly to the planned oxygen run in 2025, although they are not yet part of the official LHC plan.

%% file: intro_physics_case.tex
\section{The physics case }
\label{case}

The physics case for high-luminosity proton-nucleus collisions at the LHC is summarised in nine subsections below, each covering the following research topics: (i) Constraints on nuclear parton distribution functions, (ii)  constraints on proton GPDs and PDFs, (iii) small-$x$ QCD and gluon saturation physics, (iv) benchmark for QGP physics and onset of collectivity, (v) double and triple parton scatterings, (vi) spectroscopy of bound states, (vii) photon-photon collisions, (viii) beyond the Standard Model physics, and (ix) connections to ultra high-energy cosmic rays.

%% file: nPDF.tex
\subsection{Constraints on nuclear parton distribution functions (nPDFs and nTMDs)}
\label{sec:npdf}

\begin{figure}[h]
\begin{center}
\includegraphics[width=0.9\textwidth]{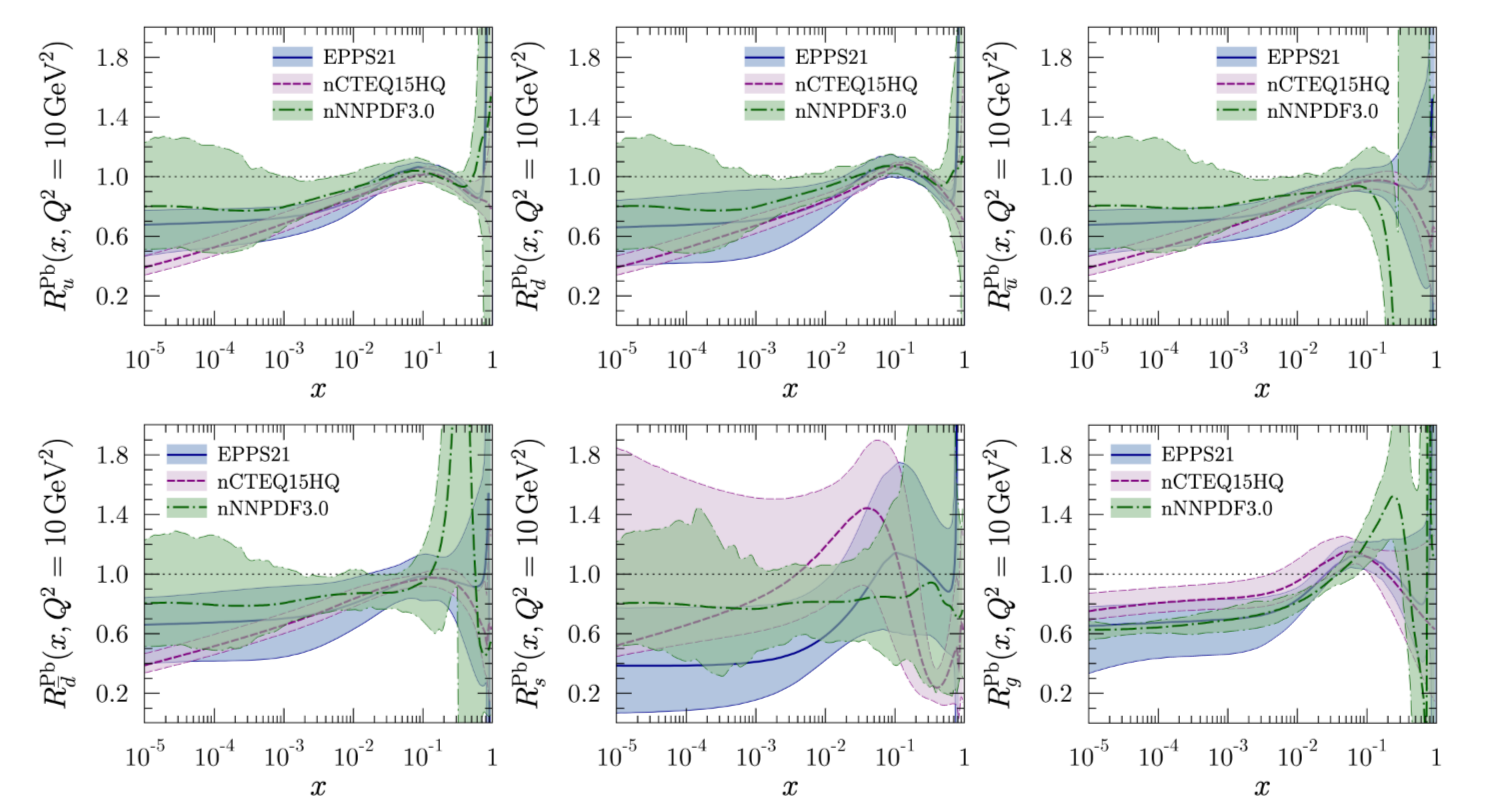}
\setlength{\belowcaptionskip}{-10pt}
\caption{Comparison of nPDF nuclear modification factors for $^{208}Pb$ (i.e.\ the parton densities of lead divided by the averaged PDFs of 82 free protons and 126 free neutrons) for different parton species from various global analyses compared in Ref.~\cite{Klasen:2023uqj}. Uncertainty bands correspond to 90\% CL. 
}
\label{fig:nPDF_errors}
\end{center}
\end{figure}

Nuclear parton distribution functions (nPDFs) describe nuclei in terms of quarks and gluons, carrying a given longitudinal momentum fraction $x$ at a factorisation scale $\mu$, and are essential universal ingredients in the description of all high-energy nuclear processes based on perturbative QCD (pQCD). When the parton density description is extended to explicitly incorporate the transverse momentum ($\kt$) of the incoming partons, they are called nuclear Transverse Momentum Dependent PDFs (nTMDs). Before LHC \pPb\ data was available, knowledge of nPDFs was relatively scarce and mostly limited to lepton DIS on fixed-target nuclei. This provided information in the rather narrow region $0.01<x<0.2$ for up and down valence quarks alone, whereas the gluon and strange nPDFs were essentially unknown and arbitrarily fixed by different nPDF parametrisations~\cite{Hirai:2007sx,deFlorian:2011fp,Stavreva:2010mw,Eskola:2009uj,Kovarik:2015cma}. 
After limited \pPb\ LHC running, the situation today is very different as summarised in Fig.~\ref{fig:nPDF_errors}.
Most of the parton densities are known with a precision better than 20\% in the region $10^{-5}<x<0.1$. This is a major improvement compared to pre-LHC times, and it needs to be highlighted that this was achieved solely thanks to \pPb\ data.
Nevertheless, the current nPDF uncertainties are still insufficient for performing precise theoretical calculations of any process involving nuclei, including heavy ions at colliders and cosmic-ray interactions. Disentangling cold nuclear-matter effects in the initial and final states, separating beyond-DGLAP parton evolution (saturation and BFKL) from other nuclear effects, or studying the QGP in PbPb collisions all depend on the baseline description obtained with nPDFs. 

A critical component for studying QCD with heavy ions 
is better information on the gluon nPDF in both the high and very low $x$ regions. Some access to high-$x$ gluons is provided by the fixed-target SMOG2 programme at the LHC~\cite{BoenteGarcia:2024kba,Hadjidakis:2018ifr}, while moderate-$x$ values can be probed at the EIC~\cite{AbdulKhalek:2021gbh}. However, low-$x$ gluons are uniquely accessible in \pPb\ collisions. The \pA\ running to date has given some constraints on the low-$x$ ($x\sim10^{-5}$) gluon distribution~\cite{Kusina:2017gkz,Kusina:2020dki,Duwentaster:2022kpv,AbdulKhalek:2022fyi,Eskola:2021nhw}. However, this information comes solely from the heavy-flavour measurements ($D$, $B$, and quarkonium production) which are also sensitive to final-state nuclear effects~\cite{Arleo:2021bpv}. 
Similar problems hold when searching for saturation or studying low-$x$ evolution, see Sec.~\ref{sec:lowx}.
To disentangle different effects, more data in the low-$x$ region are required, where a promising candidate is coherent $J/\psi$ photoproduction on the nucleus in \pA\ ultraperipheral collisions (UPCs)~\cite{Baltz:2007kq}. 

Although the photon is usually emitted from the nucleus, when it comes from the proton, its (much) higher longitudinal energy leads to differently boosted final states through which the events can be distinguished. However, the cross section is low, and sufficient statistics for PDF constraints require high luminosity. The $x\lesssim10^{-3}$ region can also be studied in \pA\ collisions through measurements of Drell--Yan (DY), isolated photons, $W$, and $Z$ boson production at forward rapidities using LHCb and ALICE detectors~\cite{Klasen:2023uqj}.
For mid- and high-$x$ gluon nPDFs, the cleanest probe is 
top-quark production, where sufficient statistics to obtain precise
differential distributions~\cite{dEnterria:2015mgr,CMS:2018qpj} require high luminosity: the current data being only sufficient to extract the total $t\bar{t}$ cross section~\cite{CMS:2017hnw,ATLAS:2024qdu}. Another option for constraining the mid-$x$ gluon is inclusive jet or dijet data, as well as $\gamma\,+\,$jet and $Z+$jet measurements~\cite{ATLAS:2018xvi}. Furthermore, the measurement of $\gamma\,+\,$heavy-quark allows probing intrinsic charm~\cite{Stavreva:2010mw}.

In addition to the constraints on the gluon nPDF, a large \pA\ dataset will allow quark nPDFs to be constrained, in particular through the aforementioned DY and $W/Z$ measurements both in central and forward/backward regions~\cite{CMS:2018qpj,ATLAS:2018xvi}. It will provide information on sea-quark PDFs, allowing for flavour separation, and access the poorl known strange nPDF. However, in this latter case much better constraints will be provided by the measurement of the rarer $W\,+\,$charm production~\cite{Stirling:2012vh}, which is directly sensitive to the strange quarks in the nucleus, and with sufficient statistics can even provide information on the $s-\bar{s}$ asymmetry.
Finally, it should be highlighted that extending the \pA\ programme to run with more than one nucleus would allow the study of the nuclear $A$-dependence of nPDFs, which is very poorly known. Currently, because of the LHC \pPb\ runs, the nPDFs of the lead nucleus are the only ones for which there are constraints for all flavours. For other nuclei which are relatively well studied, such as iron and carbon, only reliable information on up and down (anti)quarks is available. Hence, already having runs with lead and oxygen would make a big impact on our understanding of the nuclear mass dependence of nPDFs.  
{A} proton-oxygen run allowing the measurement of single differential dijet cross sections would significantly constrain gluons in light nuclei and shed light on its $A$-dependence~\cite{Paakkinen:2021jjp}. Similarly, detailed \pA\ studies are needed to probe the badly known impact-parameter dependence of nPDFs~\cite{Klein:2003dj,Helenius:2012wd,Shao:2020acd}.

With regard to studies of TMDs at the LHC, the flagship processes are the $Z$-boson transverse momentum $\pT$ spectra to constrain the quark TMD, and two-particle correlations for colour-singlet final states to study the gluon TMD distributions~\cite{Qiu:2011ai,Boer:2011kf,Sun:2011iw,Boer:2013fca,Ma:2012hh,denDunnen:2014kjo,Lansberg:2017tlc,Scarpa:2019fol}. Although the HL-LHC programme is essential for a better understanding of the gluon TMDs in the proton, through precise measurements of azimuthal correlations in \pp\ collisions~\cite{Scarpa:2019fol}, the \pA\ data would help constrain nTMDs, both for quarks and gluons. Currently, the DY process in \pPb\ collisions measured at LHC by CMS~\cite{CMS:2015zlj} and ATLAS~\cite{ATLAS:2015mwq}, together with other available lepton-nucleus data, were used to perform a global fit of the quark nTMD~\cite{Alrashed:2021csd}. In addition, experimental constraints of nTMDs will help to match the leading-power TMDs discussed here with the small-$x$ TMDs appearing in the context of gluon saturation (cf. Table~\ref{tab:smallx_TMDs} below).

%% file: 1photon.tex
\subsection{Constraints on proton GPDs and PDFs}
\label{sec:photoninduced}

UPCs at the LHC serve as an abundant source of high-energy photons making the LHC by far the most energetic photon–proton collider ever~\cite{Baltz:2007kq}. In photoproduction processes in \pA\ collisions, the ion is usually the $\gamma$ emitter, since the photon flux is proportional to the square of the ion charge, $\phi_{\gamma/A}\propto Z^2$, and this emitted photon acts as a probe of the proton.
Compared to \pp\ collisions, \pA\ collisions offer a distinct advantage when probing the nucleon since there is less ambiguity in the identity of the photon emitter (see also~\cite{Alvioli:2024cmd}). This partly compensates for the lower luminosity. A further benefit is the absence of significant pileup in \pA\ collisions in comparison to \pp\ running, where the superposition of up to 200 collisions per beam crossing in Run 4 makes the identification of photoproduction events almost impossible. Compared to \AA\ collisions in which the {\it nuclear} PDF is probed, it is the {\it free}-proton PDF that is probed in \pA\ collisions. Both exclusive and inclusive photoproduction studies are possible, and the quasi-real nature of the photon coherently emitted by the nucleus allows accurate measurements of various distributions related to the proton structure, such as PDFs and Generalised Parton Distributions~(GPDs).

Photon-induced measurements in \pA\ UPCs provide access to GPDs in a unique kinematic region different from that of fixed-target experiments (COMPASS, HERMES, JLab), and the upcoming EIC. Final states composed of two photons~\cite{Pedrak:2017cpp,Grocholski:2021man,Grocholski:2022rqj}, a meson-photon pair~\cite{Boussarie:2016qop,Duplancic:2018bum,Duplancic:2022ffo,Duplancic:2023kwe}, or a meson-meson pair~\cite{ElBeiyad:2010pji} cover the whole set of unpolarised, polarised, and transversity GPDs with small-$\xi$ reach\footnote{Here, $\xi$ is the longitudinal momentum asymmetry between the initial and final states.}, and provide complementary observables for the chiral-even sector in DVCS and DVMP. {Projections for} $p$Pb UPCs have been performed for 
$\gamma \pi^\pm$~\cite{Duplancic:2022ffo} and $\gamma \rho^{\pm,0}$~\cite{Duplancic:2023kwe}, resulting in promising anticipated statistics, both in the chiral-even and chiral-odd sectors at moderate $\xi$, the low-$\xi$ region being more favourable for the chiral-even sector. Collecting $\LumiInt = 1.2$~pb$^{-1}$ of data in Runs 3 and 4, and ensuring that the square of the $\gamma-$meson invariant mass is above 2 GeV$^2$, about $16,000\, \gamma \rho_L^0$ pairs ($1,700\, \gamma \rho_T^0$ pairs) are expected to be produced, which probe the chiral-even (chiral-odd) sector of GPDs. With the same integrated luminosity, in the small $5 \cdot 10^{-5} < \xi < 5 \cdot 10^{-3}$ range, about $800\,\gamma \rho_L^0$ pairs are expected. By extension, the study of meson-meson pair production at large invariant mass can in addition provide access to the whole set of GPDs, with the practical advantage that a photon is not required to be observed in the final state~\cite{ElBeiyad:2010pji}.

Measurements of exclusive quarkonium ($\Q$) photoproduction, $\gamma p \to \Q p$, across multiple collision systems span a broad kinematic range in $\gamma$-proton CM\ energy~\cite{ZEUS:2002wfj,ZEUS:2004yeh, H1:2005dtp, H1:2013okq, LHCb:2018rcm,LHCb:2014acg,LHCb:2015wlx}, and have been measured in \pA\ and low-luminosity \pp\ collisions at the LHC.  Such measurements are necessary for the extraction of GPDs, since these depend on both variables. These measurements extend the coverage provided by HERA and probe much higher CM\ energies than will be available at the EIC. 
The HL \pPb\ programme offers extended coverage in $x$ and greater statistical precision, particularly for $\psi(2S)$ and $\Upsilon$ measurements~\cite{Flett:2022ues}. In addition, there are reduced model dependencies in the calculation of the survival factor due to the absence of the two-fold ambiguity of the emitter~\cite{Khoze:2013dha}. The measurements will provide strong constraints on gluon GPDs~\cite{Flett:2024htj}, which are poorly known, and constrain the gluon PDFs via the Shuvaev transform for $Q^2 \approx 2.4$--$22$~GeV$^2$ and $x\approx 3\cdot 10^{-6}$--$10^{-3}$~\cite{Shuvaev:1999ce,Shuvaev:1999fm,Flett:2019pux}. Analyses using $J/\psi$ photoproduction data in \pp\ have already shown the possibility to reduce PDF uncertainties in a previously unexplored kinematic domain~\cite{Flett:2020duk, Flett:2024iex,Flett:2024uzq}. The \pPb\ measurements will decrease the theoretical uncertainties and improve the precision with which the PDFs are determined. Furthermore, measurements of higher-mass quarkonia, such as $\Upsilon$, will provide constraints on the $Q^2$ evolution in the mid-to-low $x$ domain and test factorisation~\cite{Flett:2021fvo}, while the $\sigma(\psi(2S))/\sigma(J/\psi)$ cross-section ratio will measure the structure of the radial wavefunction at the origin.
To measure the photon PDF of the proton, a new experimental method has been proposed~\cite{Luszczak:2019hao}, based on the measurement of dilepton production via $\gamma p\to \ell^+\ell^+ +X$ in \pA\ collisions. 

The $t$-dependence of $J/\psi$ photoproduction on the nucleus has been measured in PbPb collisions~\cite{ALICE:2021tyx}, but a cross-section measurement doubly differential in energy and $t$, for photoproduction on the proton, will only be possible with HL-\pA\ collisions.
A fit of the $\gamma p\to J/\psi p$ cross-section as a function of both energy and $t$ will allow a determination of the slope and intercept of the Pomeron trajectory. 
Such a measurement can also probe the validity of the widely-adopted factorisation of the $t$-dependence in PDF and GPD modelings. 
Measurements of exclusive $J/\psi$ production in \pA\ collisions can also look for higher-twist GPD contributions by testing 
$s$-channel helicity conservation through measurements of the meson polarisation.
First measurements have been performed in PbPb collisions~\cite{ALICE:2023svb}, but more data are required to perform the test in \pA\ collisions.

Besides the exclusive processes discussed above, UPCs can also lead to inclusive photoproduction processes. To date, inclusive UPCs have been only measured in PbPb collisions looking for dijet~\cite{ATLAS:2021jhn,ATLAS:2024mvt} and charm~\cite{CMS-PAS-HIN-24-003} photoproduction. Inclusive quarkonium photoproduction has been shown~\cite{Lansberg:2024zap} to be measurable by the four LHC experiments in \pPb\ UPCs, and would increase the $\pt$ reach of the HERA data~\cite{H1:1996kyo,H1:2002voc,H1:2010udv} from 10 to 20~GeV and even of the future EIC~\cite{Boer:2024ylx}.
Similarly, dijet, charm, bottom inclusive photoproduction, to name a few, will be accessible in \pPb\ UPCs and will improve the determination of the low$-x$ gluon PDFs of the proton.
In addition, the corresponding $J/\psi$, $\Upsilon$, and $\psi(2S)$ spectra, particularly at large~$\pt$, would discriminate better among different production mechanisms compared to single inclusive hadroproduction data in \pp\ at the HL-LHC. Finally, the $\pt$ spectrum of non-prompt $J/\psi$ photoproduction is also measurable, probing bottom photoproduction and improving the interpretation of the HERA $J/\psi$ data. Measurements of $J/\psi$ dissociative photoproduction are sensitive to shape fluctuations of the proton~\cite{Mantysaari:2016ykx}, in particular to energy-dependent gluonic hot spots, which should be increasingly suppressed at high energies~\cite{Cepila:2016uku}. The ALICE forward calorimeter (FoCal)~\cite{ALICE:2020mso,ALICE:2023fov} is particularly suited to directly test this phenomenon in \pPb\ collisions~\cite{Bylinkin:2022wkm}: a dissociative $J/\psi$ cross section falling with CM\ energy would signal the onset of nonlinear QCD effects.

%% file: saturation.tex
\subsection{Small-$x$ QCD and gluon-saturation physics }
\label{sec:lowx}

The phenomenon of gluon saturation arises from the nonlinear nature of QCD at high energies~\cite{Gribov:1984tu,Mueller:1985wy,Balitsky:1995ub,Kovchegov:1999yj}, and manifests itself as a breaking of the DGLAP-based description of PDFs~\cite{Altarelli:1977zs,Dokshitzer:1977sg}. It is expected that a hadronic target is found in a saturated state, when a probe (quark or gluon) scatters off a small-$x$ gluon constituent, with $x\lesssim 10^{-4}$. Due to the nonlinear effects, a dynamical saturation scale $Q_s(x)$ is generated, which is further enhanced by the target mass number, roughly $Q^2_s(x)\sim A^{1/3}$ for large nucleus~\cite{Jalilian-Marian:2005ccm,Gelis:2010nm,Albacete:2014fwa,Blaizot:2016qgz}. Therefore, direct searches for gluon saturation are best performed through scattering off heavy nuclei, and measuring forward particle production observables, using \pp\ collisions as a reference.

A suppression of the forward \pA\ cross sections (per nucleon) compared to \pp\ was observed at RHIC, both for inclusive hadron production and two-particle correlations~\cite{BRAHMS:2004xry,STAR:2006dgg,PHENIX:2011puq,STAR:2021fgw,Chu:2021hda}, and at the LHC in the inclusive hadron $\pt$ spectra measured by LHCb~\cite{LHCb:2022tjh,LHCb:2023kqs}. While qualitatively consistent with the saturation picture, the Colour Glass Condensate (CGC) theory for saturation predicts less suppression than observed~\cite{Mantysaari:2023vfh}. Indirectly, nonlinear effects may be visible~\cite{Penttala:2024hvp} in $J/\psi$ photoproduction on the nuclear target in PbPb UPCs by ALICE and CMS~\cite{ALICE:2021gpt,ALICE:2023jgu,CMS:2023snh} compared to results on the proton target~\cite{ALICE:2014eof,LHCb:2014acg,ALICE:2018oyo,LHCb:2018rcm}.
On the other hand, the data~\cite{ALICE:2021gpt} cannot distinguish between saturation~\cite{Bendova:2020hbb,Goncalves:2014wna,Lappi:2013am} and non-saturation models.
Similarly, the comparison of dijet correlations in the ATLAS forward rapidity for \pPb\ and \pp~\cite{ATLAS:2019jgo} seems to suggest a subtle interplay of nonlinear effects and perturbative Sudakov resummation~\cite{vanHameren:2019ysa}, although subject to large experimental and theoretical uncertainties. Finally, the forward inclusive jet energy deposit in \pPb\ collisions measured by CMS in the CASTOR detector~\cite{CMS:2018yhi} seems to challenge both the theoretical descriptions based on saturation~\cite{Bury:2017xwd,Mantysaari:2019nnt,Liu:2022ijp}, as well as the available Monte Carlo event generators.
 
In view of the above, a dedicated \pA\ LHC programme is crucial in disentangling the different effects and finding clear evidence for nonlinear evolution in nuclei. Since one of the essential predictions of the gluon saturation models is the collective behaviour of gluons (carrying average transverse momenta $\kt\sim Q_s$), the most significant observables are related to azimuthal particle correlations at forward rapidities. The CGC theory predicts a sensitivity of the initial-state target to the colour flow in the final state. Therefore, one of the crucial measurements are azimuthal correlations for various final states at broad transverse momentum range, including photoproduction on a nuclear target in UPCs. From a theoretical viewpoint, the simplest two-particle correlations sensitive to saturation are photon-jet and photon-hadron correlations in \pPb\ collisions at large $|\eta|$. The cross section depends on a single non-perturbative TMD small-$x$ gluon field correlator, called the dipole gluon distribution, which is also accessible in inclusive DIS~\cite{Dominguez:2011wm} (see~\cite{Jalilian-Marian:2012wwi,Rezaeian:2012wa,Benic:2022ixp,Ganguli:2023joy} for phenomenological predictions). Photon-jet correlations can be accessed up to $\eta\approx 5.1$, in the planned FoCal calorimeter of ALICE, in LHCb, and up to $|\eta| = 4$ in ATLAS and CMS at HL-LHC. In addition to the dipole TMD gluon distribution, the description of small-$x$ phenomena in the saturation regime requires other types of TMD gluon distributions~\cite{Dominguez:2011wm,Boussarie:2023izj}. The Weizs\"acker-Williams (WW)
distribution, used in the leading-power TMD factorisation formalism~\cite{Collins:2011zzd}, can be directly probed at the LHC in dijet correlations in UPC photoproduction on a nuclear target~\cite{Kotko:2017oxg,Boussarie:2014lxa,Boussarie:2016ogo}, which would complement similar future measurements performed at EIC~\cite{Iancu:2021rup}. Table~\ref{tab:smallx_TMDs} summarises the impact of the two basic small-$x$ TMD distributions to different processes accessible at HL-LHC. In particular, the precise measurement of dihadron and dijet correlations in \pPb\ collisions at forward rapidity~\cite{Jalilian-Marian:2011tvq,vanHameren:2023oiq}, as well as transverse energy-energy correlators~\cite{Kang:2024otf}, will provide stringent theory tests. 

\begin{table}[h!]
\renewcommand{\arraystretch}{1.05}
\tabcolsep=4.6mm
\caption{Impact of the two basic small-$x$ TMD gluon distributions to various processes accessible in HL proton-nucleus collisions at the LHC (adapted from~\cite{Dominguez:2011wm}). \label{tab:smallx_TMDs} }
\centering
\begin{tabular}{l|c|c|c|c}
TMD type  &  hadron in \pA\  & photon-jet in \pA\  &  dijet in $\gamma A$ (\AA\ UPC) &  dijet in \pA\  \\
\hline  WW     & \textcolor{red}{\ding{56}}    & \textcolor{red}{\ding{56}} & \textcolor{green}{\Checkmark} & \textcolor{green}{\Checkmark} \\
\hline dipole  & \textcolor{green}{\Checkmark} & \textcolor{green}{\Checkmark} & \textcolor{red}{\ding{56}} & \textcolor{green}{\Checkmark} \\
\hline
 \end{tabular}
\end{table}

Exclusive photoproduction of light mesons with a rapidity gap can help discriminate between the non-saturation (à la BFKL) and saturation (CGC) scenarios by exploiting their different $t$ dependence. Exploring the entire spin density matrix should provide a large set of observables~\cite{Enberg:2003jw, Poludniowski:2003yk,Boussarie:2016bkq,Boussarie:2024bdo,Boussarie:2024pax}, giving access to the generalised TMDs (GTMDs) of the proton. Single~\cite{Fucilla:2023mkl} and double~\cite{Fucilla:2022wcg} hadron photoproduction processes are also sensitive to GTMDs, but single inclusive observables will remain important. Last but not least, a full understanding of low-$x$ dynamics will require more direct observables of the intermediate region of high gluon density targets, where the low-$x$ linear BFKL energy evolution is needed, but saturation is not relevant yet.  In particular, despite there being indirect experimental evidence for the Odderon in elastic \pp\ scattering~\cite{D0:2020tig}, there are no experimental hints of the Odderon in the hard sector. To address this, \pA\ collisions provide good prospects for observation through interference effects in exclusive $\pi^+\pi^-$ photoproduction~\cite{Hagler:2002nf}; by the observation of $C=+1$ mesons in photoproduction~\cite{McNulty:2020ktd}; and through the {transverse} momentum distribution of exclusive $J/\psi$ mesons~\cite{Bzdak:2007cz}. 

%% file: AA-benchmark.tex
\subsection{Benchmark for QGP physics and onset of parton collectivity}

One of the main motivations for studying \pA\ collisions at the LHC was to obtain a reliable baseline, without final-state effects, to interpret the \AA-collision results. However, with the rising interest in parton collectivity in small systems (e.g.\ in the creation of QGP in \pPb\ collisions) the \pA\ programme itself merits dedicated study.
Historically, \pA\ collisions at the LHC and RHIC were mainly motivated by measuring the so-called ``cold'' nuclear matter effects on strongly interacting probes of the QGP. Their main purpose was to measure how the production of such probes was suppressed by the modification of the nuclear partonic densities, by initial-state energy loss, and/or by final-state interactions~\cite{Accardi:2009qv}. The \pA\ runs have proven to be absolutely essential as they uncovered a variety of unexpected effects which need to be understood in their own right, as well as being crucial for the interpretation of \AA\ measurements. 
In particular, the observations of azimuthal correlations that are long-range in rapidity, are indicative of collective behaviour in high multiplicity \pp~\cite{CMS:2010ifv} and \pPb~\cite{ALICE:2012eyl,ATLAS:2012cix,CMS:2015yux} collisions.
This observed collectivity in small systems has triggered intensive research~\cite{Noronha:2024dtq,Grosse-Oetringhaus:2024bwr} to understand its origins. Explanations range from strong final-state effects similar to \AA\ collisions, to initial-state effects due to gluon saturation, while the success of hydrodynamic models in describing the \pPb\ data calls for further research. High luminosity \pPb\ collisions also provide a unique opportunity to study complex vortex-like structures in QGP droplets; for example, studies of hyperon polarisation could lead to the discovery of the toroidal vorticity in nuclear matter~\cite{Lisa:2021zkj,DobrigkeitChinellato:2024xph}.
Therefore, while more detailed \pA\ studies of cold nuclear matter effects, e.g.\ the relative suppression of excited quarkonia compared  to their ground states in \pPb\ collisions, are needed, a HL-\pA\ run at the LHC will also provide much further information on the origins of parton collectivity.

%% file: DPS-TPS.tex
\subsection{Double and triple parton scatterings} 

Double and triple parton scattering (DPS and TPS) processes in high-energy hadron-hadron collisions open up novel opportunities to investigate the partonic hadron structure~\cite{Diehl:2011yj,Gaunt:2009re,dEnterria:2016ids,Rinaldi:2018slz}, and complement the multidimensional picture of hadrons as described by GPDs and TMDs. In addition, DPS and TPS final states constitute backgrounds for BSM searches (see e.g. Ref.~\cite{CMS:2022pio}). Although the DPS and TPS signals are typically much smaller than the equivalent signal produced in single parton scattering (SPS) processes, they can be enhanced by extending the transverse size of one of the colliding hadrons using heavy nuclear targets\footnote{In \pPb\ collisions ($A=208$), DPS and TPS yields are enhanced by factors of about $3\times A$ and $9\times A$ compared to \pp\ collisions~\cite{dEnterria:2017yhd}.}~\cite{Strikman:2001gz,Cattaruzza:2004qb,dEnterria:2012jam,Blok:2012jr,dEnterria:2014lwk}, whereas they are completely swamped by binary-scaling contributions from different nucleon-nucleon scatterings in \AA\ collisions~\cite{dEnterria:2013mrp,dEnterria:2017yhd}. Multiple experimental analyses of DPS have been performed, whose results are usually summarised through the extraction of the so-called effective cross section, $\sigma_\text{eff}$, defined as the normalised ratio of SPS to DPS cross sections for the same final states. This quantity provides critical insights into the transverse hadron structure, and badly known double parton correlations~\cite{Rinaldi:2018slz}. In a purely geometric approach, $\sigma_\text{eff}$ is assumed to be a process-independent constant~\cite{dEnterria:2017yhd,Blok:2017alw}, although recent compilation of measurements show differences between $\sigma_\text{eff}$ extracted from processes involving quarkonium~\cite{CMS:2021qsn,Lansberg:2019adr,Lansberg:2017chq,Lansberg:2016rcx,ATLAS:2016ydt,Lansberg:2014swa} and jets or gauge-boson production~\cite{CMS:2024wgu}. Parton correlations might explain these discrepancies, which can be better investigated with HL-\pPb\ data.

Estimates for DPS and TPS contributions to heavy-quark, quarkonium and/or electroweak pair production in \pPb\ collisions at the LHC have been provided in Refs.~\cite{dEnterria:2014lwk,dEnterria:2018cvh}. To date, only two experimental extractions of $\sigma_\text{eff}$ exist in \pPb\ collisions from double charm mesons~\cite{LHCb:2020jse} and double $J/\psi$~\cite{CMS:2024wgu} production. Both measurements are statistically limited and more data are required. A clean extraction of $\sigma_\text{eff}$ is possible from same-sign W boson production~\cite{dEnterria:2012jam}, where a few pb$^{-1}$ of data would allow a precision of 10\%. A comprehensive investigation of DPS in gluon-initiated processes is possible through $J/\psi+\Upsilon$ production~\cite{Chapon:2020heu}. Currently, only a limited number of events have been observed by CMS in \pp\ collisions, and the use of Pb nuclei would increase the corresponding rate. Moreover, double-$\Upsilon$ production would enable a comparative analysis of the extracted $\sigma_\text{eff}$ with that obtained in double-$J/\psi$ production, providing insights into the role of final-state interactions. Finally, large data samples are needed to carry out multidifferential studies of $\sigma_\text{eff}$, e.g.\ as a function of difference in rapidities or azimuthal angles between final states~\cite{Cotogno:2020iio,Ceccopieri:2017oqe}. These analyses will provide unique information on double-parton correlations that cannot be accessed otherwise. The separation of DPS from SPS processes can also be facilitated in \pA\ collisions by exploiting their different centrality dependence~\cite{Alvioli:2019kcy}.

A more detailed understanding of DPS can be achieved through the study of TPS in \pp~\cite{dEnterria:2017yhd} and \pPb~\cite{dEnterria:2016yhy} collisions. In \pp, TPS has been searched for in triple $J/\psi$ production~\cite{CMS:2021qsn}, and also in this case, the rate will be enhanced in \pPb\ collisions. Large data samples offer unique opportunities to observe TPS, e.g., in $\phi \phi D$ or $\phi \phi J/\psi$ production. Note that in \pp\ (\pPb) collisions, triple charm production is almost $15\,(20)\%$ of all inclusive charm production, so not only the role of TPS cannot be neglected, but it is mandatory to properly characterise the corresponding final states~\cite{dEnterria:2017yhd}. Another promising channel is 6-jet production, where the impact of TPS can be  up to $20\%$ of the total cross section above $\pt^\text{jet}\approx 20$~GeV in \pp\ collisions, an effect that is further increased in the \pPb\ case~\cite{Maneyro:2024twb}. 

%% file: spectroscopy.tex
\subsection{Production and spectroscopy of bound states}

In the quarkonium sector, LHCb has measured multiple states in \pPb\ collisions~\cite{LHCb:2023apa}, covering a broad range of binding energies and sizes. After accounting for initial-state effects, the data reveal a trend of dissociation of quarkonium states with weak binding, such as $\psi(2S)$, and production consistent with scaled \pp\ collisions for states with binding energy larger than 180~MeV. The exceptions are prompt $\Upsilon(2S)$ and $\Upsilon(3S)$ states, which show anomalous suppression relative to the $\Upsilon$(1S) yields~\cite{LHCb:2018psc}. The potential cause of these suppressions is the feed-down contribution of weakly bound $\chi_b$ states~\cite{LHCb:2014ngh}. However, $\chi_b$ states were never measured in \pPb\ collisions due to the low efficiency for low-energy photons produced in the decay $\chi_{b}\to \Upsilon + \gamma$, and more data are required. Multiplicity-dependent measurements of $\chi_{c}\to J/\psi+\gamma$ states to search for anomalous suppression in high-density events, as well as measurements of the $\chi_{b}\to \Upsilon+\gamma$  feed-down contributions are essential to confirm the origins of $\Upsilon(2S)$ and $\Upsilon(3S)$ anomalous suppression. The challenging observation of the $\eta_c$ state will only be possible at high luminosity and will allow a better understanding of the behaviour of colour-singlet states in the nuclear medium.

In the field of exotic hadrons, 75 new hadrons have been discovered at the LHC to date~\cite{LHCb-FIGURE-2021-001-report} including many combinations of (candidate) tetraquark and pentaquark states. The nature of a large number of these exotic hadrons is still under debate as it is not clear whether they are tightly bound or molecular-like. Their binding and configuration are still largely unknown, and so $pA$ collisions function as an excellent laboratory to study their properties. Of particular interest is to understand how these exotic states are produced and interact in high-density environments. LHCb observed an enhancement of tetraquark $\chi_c$(3872) production in \pPb\ compared to \pp\ collisions~\cite{LHCb:2024bpb}, hinting at the role of statistical hadronisation in their formation, where the larger number of initial-state quarks increase the probability of multiquark hadron production~\cite{ExHIC:2017smd}. High-luminosity \pPb\ collisions can provide more precise measurements of $\chi_c$(3872) and other exotic hadrons, including their density-dependent production and ``destruction''. One expects a trade-off between statistical-hadronisation formation and dissociation of weakly-bound exotic states by comoving particles~\cite{Esposito:2020ywk}, depending on their microscopic nature. Exotic production from intrinsic charm~\cite{Vogt:2024fky} can also be searched for in HL-\pPb\ collisions, where the asymmetric collision can isolate charm-rich partonic environments at large $x$.

%% file: 2photon.tex
\subsection{Photon-photon collisions}
\label{sec:2photon}

Both LHC protons and heavy ions can act as sources of initial-state photons and hence photon-photon collisions occur abundantly in UPCs at the LHC~\cite{Baltz:2007kq}. Being a colour-singlet exchange, $\gamma\gamma$ collisions naturally lead to events with intact projectiles and rapidity gaps in the final state. Together with low-pileup conditions, UPCs give very clean experimental signatures with very few particles registered in a detector. The photon flux accompanying each beam is proportional to $Z^2$, thus, cross sections for $\gamma\gamma$ processes are significantly enhanced in \AA\ compared to \pA\ and \pp\ collisions. 
While the $\gamma\gamma$ luminosities in \pA\ collisions are overall reduced by a factor $Z^2$ compared with the \AA\ case, the proton beam energies are larger, and the associated photon fluxes are much harder, than in \PbPb\ UPCs. As a result, the \pA\ collisions probe significantly larger $\gamma\gamma$ CM\ energies~\cite{dEnterria:2009cwl}, and they are also useful to resolve discrepancies between \pp\ and \AA\ UPCs. Already some hints of mild deviations~\cite{ATLAS:2020epq,ATLAS:2022srr,ATLAS:2020mve} between data and LO predictions exist for exclusive $e^+e^-$ and $\mu^+\mu^-$ production, that highlight the need for a proper modeling of inelastic contributions~\cite{Harland-Lang:2020veo} as well as of the Pb photon flux and higher order QED corrections~\cite{Shao:2024dmk}. Additional datasets with the asymmetric $\gamma\gamma$ collisions provided by \pA\ UPCs can help clarify all these aspects. Also, particular UPC processes possible in the \pA\ mode, such as single-$W$ photoproduction~\cite{Dreyer:2009zz}, require large data samples. Forward neutron production from electromagnetic ion dissociation has gained interest in \AA\ UPCs~\cite{ALICE:2012aa,ATLAS:2020epq,CMS:2020skx,ALICE:2022iqi,ATLAS:2022srr,CMS:2024bnt} and is increasingly used in online and offline event selection of SM processes, as well as in BSM searches~\cite{ATLAS:2024nzp}. 
Different neutron multiplicities have different impact-parameter profiles that lead to modifications of central kinematics. The simplicity of dilepton production in the \pA\ system, which constrains neutron emission from just one nucleus, will improve the modeling of dissociation for the \AA\ system. 

%% file: BSM.tex
\subsection{Beyond the Standard Model}

At face value, \pA\ cannot compete with \pp\ collisions at the LHC in terms of the production of heavy BSM objects, as they have lower CM\ energies and integrated luminosities. However, akin to the \AA\ case~\cite{Bruce:2018yzs,dEnterria:2022sut}, \pA\ collisions feature $\gamma\gamma$ interactions without pileup and with large photon fluxes (from the Pb side) that partially compensate for these drawbacks (provided that a large $\LumiInt$ is warranted) for photon-coupled BSM objects. In terms of attainable $\gamma\gamma$ luminosity, the HL-\pPb\ mode would outperform the \PbPb\ UPCs reach in the $m_{\gamma\gamma}\approx 50$--300~GeV mass range~\cite{dEnterria:2009cwl}. This is relevant e.g.\ to set competitive limits on heavy axion-like particles~\cite{Knapen:2016moh}.

In order to compare the generic BSM reach of \pA\ compared with \pp\ and \AA\ collisions, we introduce a simple ansatz for the $\gamma\gamma$ collision cross section: 
\mbox{$    
\sigma_{\gamma\gamma}(n)\propto  {s_{\gamma\gamma}^{n-1}}/{\Lambda^{2n}}
$}. This simplified cross section encodes the CM energy ($\sqrt{s_{\gamma\gamma}}$) dependence, which is one of the key elements to compare  the collision modes.  
In a specific model,  the $\Lambda$ parameter would generally encapsulate a combination of couplings and masses. 
This simplified approach provides a rough classification of BSM candidates as a function of $n$.   The value 
$n=0$ includes SM-like processes 
(see e.g.~\cite{Fichet:2014uka}), $n=1$ includes resonant effective field theories (EFTs) (see e.g.~\cite{Fichet:2015yia,Baldenegro:2022kaa}), and
$n\geq2$ arises from non-resonant EFTs, such as $F^4$ operators~\cite{Fichet:2013ola} and continuum EFTs~\cite{Megias:2019vdb,Fichet:2022ixi}. Using this approach, we classify the BSM scenarios to be searched in UPCs into (roughly) two types: low-mass resonances and non-resonant EFTs.

The low-mass resonances\footnote{UV motivated CP-odd resonances include the PQ axion~\cite{Peccei:1977hh,Peccei_PRL}, stringy axions~\cite{Witten:1984dg, Conlon:2006tq, Svrcek:2006yi,Arvanitaki:2009fg,Acharya:2010zx,Cicoli:2012sz}, and Goldstone bosons~\cite{Masso:1995tw}, whereas CP-even resonances include the radion~\cite{Csaki:2000zn}, dilaton~\cite{Goldberger:2007zk}, composite radial mode~\cite{Fichet:2016xvs,Fichet:2016clq}, extended Higgs sectors~\cite{Gunion:1989we}, Higgs portal~\cite{Englert:2011yb}, and {KK gravitons~\cite{Csaki:2004ay,Fichet:2013ola}  }.} are described by resonant EFTs.
In that case, we obtain  that the \pPb\ mode competes with \pp\ mode. The non-resonant EFTs include anomalous quartic gauge couplings\footnote{UV motivation includes heavy neutral particles linearly coupled to the SM~\cite{Fichet:2013ola,Baldenegro:2017aen}, new charged particles~\cite{Fichet:2014uka,Baldenegro:2017aen}, polarisable dark particles~\cite{Fichet:2016clq} and Born-Infeld QED~\cite{Gibbons:1997xz,Davila:2013wba,Ellis:2017edi}.} and continuum EFTs\footnote{UV motivation includes AdS~\cite{Randall:1999vf,Strassler:2008bv,Friedland:2009iy,Friedland:2009zg,Chaffey:2023xmz}, 
linear dilaton~\cite{Megias:2019vdb,Megias:2021arn,Fichet:2022ixi,Fichet:2023xbu}  and other braneworld geometries~\cite{Fichet:2023dju}, and strongly-interacting dark sectors~\cite{McDonald:2010iq, vonHarling:2012sz, Brax:2019koq}.}. In that case, we obtain  that \pPb\ competes with \pp\ with an event yield only moderately smaller, but with a much cleaner selection due to reduced pileup. Due to this complementarity, assuming no prejudice on the type of BSM scenario, searches using HL-\pPb\ collisions provide a useful strategy to maintain sensitivity over the broadest range of  new physics possibilities. More detailed studies of effective operator searches in \pPb\ would be useful to reinforce such a conclusion.

%% file: UHECR.tex
\subsection{Ultra high-energy cosmic-ray physics}
\label{sec:cosmic}

Cosmic rays, ranging from medium to ultra-high energies, originate from various astrophysical sources and produce extensive air showers (EAS) upon interaction with atmospheric nuclei. These showers provide valuable information about the primary particles, including their mass composition and energy spectrum. The study of cosmic rays and EAS offers a unique opportunity to probe the behaviour of strongly interacting matter under extreme conditions, provided that the hadronic interaction models reproduce the \pA\ collider data up to the highest possible energies~\cite{dEnterria:2011twh,Kampert:2012mx}. The analysis of EAS has highlighted challenges, notably the ``muon puzzle''~\cite{PierreAuger:2020gxz,PierreAuger:2021qsd}, whereby current hadronic interaction models fail to accurately predict muon production for a given primary mass~\cite{Cazon:2020zhx}. An improved description of the EAS data requires significant changes to the models, both for their predicted position of the shower maximum and for the fraction of signal at ground associated to the number of muons~\cite{PierreAuger:2024neu}. This discrepancy suggests an incomplete understanding of hadronic interactions at ultrahigh energies and has sparked interest in exploring new phenomena, in particular linked to nuclear effects in small systems~\cite{Albrecht:2021cxw}. As a matter of fact, one of the sources of the ``muon puzzle'' is seemingly the presence of \pA\ collisions that do not behave as a simple superposition of \pp\ interactions~\cite{Pierog:2023ahq}. Similarly, another area where more LHC \pA\ data are welcomed is in the interpretation of ultrahigh-energy astrophysical neutrinos as measured by the IceCube experiment~\cite{IceCube:2013low,IceCube:2014stg}. A background to cosmic neutrinos comes from atmospheric neutrinos produced in cosmic-ray interactions with air nuclei. The HL-\pA\ data can provide more precise information on the production of forward charmed particles, which are important to constrain this background~\cite{Goncalves:2017lvq}.

%% file: summary-conclusions.tex
\section{Summary and conclusions}

High-energy proton-nucleus collisions (\pA) provide a bridge between proton-proton (\pp) and ion-ion (\AA) collisions with two particular merits: firstly, the asymmetric projectiles and beam energies ensure that effects associated with the proton can be distinguished from those of the nucleus; secondly, a path towards understanding the complexity of two large systems of colliding bound nucleons is provided through interaction of a well-understood proton on the complex ion.

At the LHC, a few weeks of \pA\ collisions were performed in Runs 1 and 2 in a diversity of operating conditions, and have already brought essential contributions to particle, heavy-ion, and cosmic-ray physics, leading to the discovery of new phenomena as well as the confirmation and extension of effects discovered in lepton-nucleus collisions. Multi-TeV \pA\ collisions offer several unique physics opportunities:
\begin{itemize}\setlength{\itemsep}{0.cm}
\item {They provide a way to study nuclear modifications to PDFs and cold nuclear matter effects (such as shadowing, parton saturation, and energy loss) without the complexities of hot QCD medium effects present in \AA\ collisions.}
\item {They serve as a crucial reference for disentangling initial-state nuclear effects from final-state medium effects in \AA\ collisions, aiding in the interpretation of quark-gluon plasma (QGP) signatures.}
\item {They typically achieve higher luminosities than \AA\ collisions and much reduced pileup compared to \pp\ collisions, enabling more precise measurements.} 
\item {They provide a unique platform to extend studies of small-$x$ QCD and gluon saturation by probing smaller momentum fractions than any other current experimental setup.} 
\item {They offer valuable insights
into the interplay of enhancement and suppression mechanisms in nuclear matter and a cleaner environment to study coalescence and fragmentation in hadronisation. } 
\item {They are essential for the modelling of high-energy cosmic-ray and neutrino interactions, providing a link between particle physics and astrophysics.} 
\end{itemize}

The physics potential provided by the proton and heavy-ion LHC beams in asymmetric \pA\ collisions offers multiple complementarities and advantages compared to \pp\ and \AA\ collisions. We have summarised the physics case for a high-luminosity \pA\ run (HL-\pA@LHC) under twelve research axes, where large data samples are required to reduce the current experimental uncertainties and/or to study (for the first time) multiple rare processes of interest.
Table~\ref{tab:comp_qual} gathers qualitative comparisons of the impact that \pA\ can have on each of these physics topics, with respect to other collision scenarios: EIC~\cite{AbdulKhalek:2021gbh}, FT@LHC~\cite{Hadjidakis:2018ifr}, \pp@HL-LHC~\cite{Azzi:2019yne}, \AA@HL-LHC~\cite{Citron:2018lsq}; with $\star$ symbols assigned as explained below.\\

\begin{table}[h]\renewcommand{\arraystretch}{1.05}
\caption{Qualitative comparison of various experimental setups with respect to different physics observables. An increasing number of star points indicates a better environment for the considered physics topic.}
\label{tab:comp_qual}
\centering
\resizebox{0.9\textwidth}{!}{
\begin{tabular}{p{4.5cm}|c|c|c|c|c} Physics topics \textbackslash~Collider   & EIC & FT@LHC & HL-\pA@LHC & \pp@HL-LHC & \AA@HL-LHC \\ \hline 
PDFs & $\star\star\star\star$ & $\star\star\star$ & $\star\star$ & $\star\star\star$ &  -- \\ \hline
nPDFs & $\star\star\star\star$ & $\star\star\star$  & $\star\star\star$ & -- & $\star\star$ \\ \hline
TMDs & $\star\star\star$ & $\star\star\star$ & $\star\star\star$ & $\star\star$ & -- \\ \hline
nTMDs & $\star\star\star$ & $\star\star$ & $\star\star$ & -- & $\star$ \\ \hline
GPDs & $\star\star\star$ & $\star$  & $\star\star$ & $\star$ & -- \\ \hline
nGPDs & $\star\star\star$ & $\star$  & $\star$ & -- & $\star\star$ \\ \hline
Parton saturation searches &  $\star\star\star$ & $\star$ & $\star\star\star\star$ &  $\star\star\star$ & $\star\star\star$ \\ \hline
Odderon searches &  $\star\star$ & $\star$ & $\star\star\star$ &  $\star\star$ & $\star\star$ \\ \hline
Parton collectivity & -- & $\star\star\star$ & $\star\star\star$ & $\star\star$ & $\star\star\star\star$ \\
 \hline
DPS/TPS & $\star$  & $\star$ & $\star\star\star\star$ &  $\star\star\star\star$ & $\star$ \\
\hline
Hadron spectroscopy & $\star$ & $\star$ & $\star\star\star$ & $\star\star\star$ &  $\star\star$ 
 \\ \hline
BSM searches 
& $\star$ & $\star$ & $\star\star$ & $\star\star\star\star$ & $\star\star$ 
 \\ \hline
\end{tabular}
}
\end{table}

Regarding 
\begin{itemize}
    \item PDF studies, the best experimental setup is given by the future EIC with cleaner probes and access to polarised PDFs, while \pp@HL-LHC and FT@LHC complement its reach by probing the low-$x$ and large-$x$ regimes, especially in the gluon sector. HL-\pA@LHC can contribute in a timely manner to gluon PDF studies via inclusive photoproduction ($Q^2 \simeq 0$) in UPCs with a significantly wider range in $\pT$ and $\gamma$-proton CM energy
    than at HERA, and with indirect PDF constraints via exclusive $\Q$ photoproduction.
    \item nPDF studies, the best setup is the EIC as an $eA$ collider with cleaner final states. However, HL-\pA@LHC covers much lower $x$, while FT@LHC offers access to larger $x$ with more versatility in the probed nuclei, and earlier than the EIC. The \AA@HL-LHC programme offers some sensitivity on nPDFs if the additional hot nuclear effects can be separated out.
    \item TMD studies, the best setups are the future EIC, FT@LHC, and HL-\pA@LHC for very different reasons. On the one hand, FT@LHC would benefit from the possibility to polarise the target for single transverse-spin asymmetries studies of gluon-sensitive probes, which are essentially unknown. The EIC will profit from, both, beams and target polarisations and from cleaner probes. On the other hand, HL-\pA@LHC can study TMDs in the low-$x$ regime in the dense-dilute limit, which cannot be accessed otherwise, as well as potentially via azimuthal asymmetries in inclusive photoproduction. \pp@HL-LHC can also definitely help in probing TMDs at low $x$ via azimuthal asymmetries but kinematical cuts due to triggers are usually very harmful. 
    \item nTMD studies, the best setup is the future EIC as an $eA$ collider that can access a large variety of TMD-factorisable processes. The FT@LHC and HL-\pA@LHC programmes can access nTMDs via DY final states. Whereas \AA@HL-LHC can in principle measure nTMDs via azimuthal asymmetries in photoproduction (UPCs), the event counts are expected to be very small.
    \item GPD studies, the best setup is the EIC with polarised beams, while HL-\pA@LHC via UPCs offers very interesting possibilities through exclusive-photoproduction channels like timelike Compton scattering, or with large final-state invariant-mass systems in meson-pair or photon-meson photoproduction. Similar exclusive final states can be studied at \pp@HL-LHC but would be ``polluted" by hadronic exchanges.
    \item nGPD studies, the best setup is the EIC via $eA$ exclusive reactions which can be complemented at \AA@HL-LHC by exclusive-$\Q$ photoproduction reactions via UPCs.
HL-\pA@LHC has a limited sensitivity to nGPDs with the same observable in the rapidity region where the probability for photon emission by the proton becomes significant.
Finally, at FT@LHC in PbA UPCs, the nGPDs of various nuclear targets could also be probed via exclusive-$\Q$ photoproduction. 

    \item parton-saturation searches, the best setup is the HL-\pA@LHC via forward hadron and jet production processes and their correlations. At \AA@HL-LHC, saturation can be studied via several UPC observables while \pp@HL-LHC is needed both as a reference for saturation studies in nuclei and to test the need for the small-$x$ resummation. While the future EIC  will study inclusive and exclusive processes sensitive to saturation  in $eA$ collisions, the  forward detector upgrades of the LHC experiments will allow probing the partonic structure of heavy nuclei at much smaller $x$. Diffractive processes in UPCs are particularly promising since their extended kinematics offers a bridge towards GTMD studies.
    \item odderon searches, the best setup is the HL-\pA@LHC through the observation of $C=+1$ mesons (or meson pairs, e.g.\ $\pi^+ \pi^-$) photoproduction, which is sensitive to interferences of $C=+1$ (pomeron) and $C=-1$ (odderon) exchanges. Such processes are also possible at the EIC, but require very high luminosity. Good prospects are also presented in UPCs at \AA@HL-LHC if the photon emitter can be identified.
    \item parton-collectivity studies, the best setup is the \AA@HL-LHC as a laboratory for QGP creation. FT@LHC via Pb$A$ collisions can study collectivity in a complementary rapidity and (lower) energy domain. However, HL-\pA@LHC, and to a lesser extent \pp@HL-LHC,  can provide new probes of the generation of collective partonic behaviour in small systems.
    \item DPS/TPS studies, the best setups are HL-\pA@LHC (thanks to the $3 \times A$ and $9\times A$ enhanced yields, respectively, for \pPb\ compared to \pp\ collisions) and \pp@HL-LHC (thanks to its very large integrated luminosity and higher $\sqrts$, enabling the production of pairs of very heavy particles) with large rates for multiple DPS/TPS processes. These provide access to novel information on the partonic structure that cannot be obtained elsewhere. HL-\pA@LHC will facilitate the determination of the effective cross section as a function of multiple kinematic variables, hence revealing previously unexplored multiparton correlations.
    \item hadron spectroscopy, the best setups are HL-\pA@LHC and \pp@HL-LHC. The latter has allowed the identification of a great number of exotic states, while the former is important in elucidating their nature by exploiting their density-dependent production and final-state interactions.
    \item BSM searches, the best setup is \pp@HL-LHC, while both HL-\pA@LHC and \AA@HL-LHC  UPCs provide a clean environment for low- and intermediate-mass photon-coupled BSM objects, such as new even-spin particles. In  UPC searches for low-mass resonances, \pPb\ competes with PbPb in sensitivity, whereas in UPC searches for non-resonant EFTs,  \pPb\ has slightly lower yield than \pp\ but cleaner selection due to reduced pileup.    
    Due to this complementarity, assuming no prejudice on the type of the BSM scenario, searches using HL-\pPb\ collisions can provide a  strategy to maintain sensitivity over the broadest range of  new physics possibilities.
    \end{itemize} 

It appears clear from the above that the allocation of dedicated \pA-collision run(s) at the LHC will provide unique physics inputs complementary to those of other major existing or planned facilities. We therefore strongly encourage additional \pA\ running at the LHC in Runs 3 and 4, in order to achieve and extend the original physics targets in a timely manner, and to provide the data for the many important measurements summarised in this document.

%% file: acknowledgements.tex
\section*{Acknowledgements and funding}

This work has received funding from the French CNRS via the IN2P3 projects ``QCDFactorisationAtNLO'' and the COPIN-IN2P3 project \#12-147 ``kT factorisation and quarkonium production in the LHC era'', via the ANR under the grant ANR-20-CE31-0015 (``PrecisOnium''), by the Paris-Saclay U. via the P2I Department and by the GLUODYNAMICS project funded by the ``P2IO LabEx (ANR-10-LABX-0038)" in the framework ``Investissements d’Avenir" (ANR-11-IDEX-0003-01) managed by the Agence Nationale de la Recherche (ANR), France.
The material presented has been based upon work supported also by the U.S. Department of Energy, Office of Science, Office of Nuclear Physics, under DOE Contract No.~DE-SC0012704 and within the framework of the Saturated Glue (SURGE) Topical Theory Collaboration.
The research conducted in this publication was funded by the Irish Research Council under grant number GOIPG/2022/478 and the Joint PhD Programme of Universit\'e Paris-Saclay 1248 (ADI).
A.K. acknowledges the support of Narodowe Centrum Nauki under Sonata Bis Grant No. 2019/34/E/ST2/00186.
I.G-B was supported by program ``Excellence initiative – research university'' for the AGH University of Krakow, grant no 9722.
C.V.H. has received funding from the programme ``Atracci\'on de Talento'', Comunidad de Madrid (Spain), under the grant agreement No 2020-T1/TIC-20295.
LHL thanks the Science and Technology Facilities Council (STFC) part of U.K. Research and Innovation for support via the grant award ST/T000856/1.
F.J. has been supported by the US Department of Energy.
The work of L.S. was supported by the Grant No. 2024/53/B/ST2/00968 of the National Science Centre.
The research of M.S. was supported by the US Department of Energy Oﬃce of Science, Oﬃce of Nuclear Physics under Award No. DE-FG02-93ER40771.